\documentclass[12pt]{article}
\pdfoutput=1
\usepackage{graphicx,amssymb,amsfonts,amsmath,amssymb,amscd,amstext, mathrsfs}
\makeatletter

\@addtoreset{equation}{section}
\newcommand{\be}{\begin{eqnarray}}
\newcommand{\ee}{\end{eqnarray}}
\newcommand{\ba}{\begin{array}}
\newcommand{\ea}{\end{array}}
\newcommand{\nn}{\nonumber}

\makeatletter \@addtoreset{equation}{section} \makeatother

\begin{document}
~
\vspace{0.5cm}
\begin{center} {\Large \bf  Holographic Butterfly Velocities in Brane Geometry\\ 
\vspace{0.2cm}
and Einstein-Gauss-Bonnet Gravity with Matters}                                                  
\vspace{1cm}

                      Wung-Hong Huang \\
                      \vspace{.2cm}
                      Department of Physics\\
                       National Cheng Kung University\\
                       Tainan, Taiwan\\                      

\end{center}
\vspace{1cm}
\begin{center} 
{\large \bf  Abstract} 
\end{center}

In the first part of the paper we generalize the butterfly velocity formula to anisotropic spacetime.  
We apply the formula to evaluate the butterfly velocities in M-branes, D-branes and strings backgrounds.  We show that the butterfly velocities in M2-branes, M5-branes and the intersection  M2$\bot$M5  equal to those in fundamental strings, D4-branes and the intersection  F1$\bot$D4 backgrounds, respectively.   These observations lead us to conjecture that the butterfly velocity is generally invariant under a double-dimensional reduction.   
In the second part of the paper, we study the butterfly velocity for Einstein-Gauss-Bonnet gravity with arbitrary matter fields.  A general formula is obtained.  We use this formula to compute the butterfly velocities in different backgrounds and discuss the associated properties.

\thispagestyle{empty}

\vspace{2cm}
\begin{flushleft}
*E-mail:  whhwung@mail.ncku.edu.tw
\end{flushleft}
%%%%%%%%%%%%%%%%%%%%%%%%%%%%%%%
\newpage
%%%%%%%%%%%%%%%%%%%%%%%
\tableofcontents
\newpage
%%%%%%%%%%%%%%%%%%%%%%%
\section {Introduction}

Quantum chaos is naturally characterized by the commutator  $[W(t,x),V(0)]$ which measures the dependence of a later operator $W(t,x)$ on an earlier perturbation $V(0)$. The strength of the associated butterfly effect can be described by [1,2,3]
\be
\langle [W(t,x),V(0)]^2\rangle_T \sim e^{\lambda(t-t_*-{|x|\over v_B})}
\ee
where $t_*$ is a time scale called the scrambling time at which the commutator grows to be ${\cal O}(1)$. 
The buttery velocity $v_B$ characterizes speed at which the perturbation grows.  
The Lyapunov exponent $\lambda$ measures the rate of growth of chaos, and it is bounded by temperature T: $\lambda \le 2\pi\beta$ where $\beta={1\over T}$.
The inequality saturates for the thermal systems that have a dual holographic black hole described by Einstein gravity [4]. 

Butterfly effects in quantum chaos have been extensively studied recently in the holographic theories [5-26]. 
In the holographic approach the butterfly velocity is identified by the velocity of shock wave which describes how the perturbation spreads in space [8,10,11].  
The method of finding the shock wave  velocity for the general spacetime with matters had been described in many years ago [27], and was used to obtain the butterfly velocity.
In this paper we consider futher generalizations. 
In the first part of this paper we extend the known butterfly velocity formula to the anisotropic spacetime. We will apply our formula to several brane systems. 
Our computations lead us to conjecture that the butterfly velocity is a quantity that is invariant under a double-dimensional reduction.
In the second part of the paper, we consider the butterfly velocity for the Gauss-Bonnet gravity with arbitrary matter fields. Butterfly velocity in the higher-derivative gravity including Gauss-Bonnet term without matter fields had been discussed in [8] and [18].
The Gauss-Bonnet gravity is a simplest correction of the Einstein theory without introducing derivatives higher than second appearing in the field equation. 
In the AdS/CFT correspondence, the Gauss-Bonnet term in the bulk corresponds to next-to-leading order corrections in the 1/N expansion of the dual CFT [28].
Here we add the matter fields and apply our general formula to evaluate the butterfly velocity in several  interesting holographic systems. 

After reviewing the previous method, we extend the butterfly velocity formula [8,10,11] to the anisotropic spacetime in section 2.   
We present a rather simple way, i.e., without too many complicated tensor calculations, to derive a general formula given in \eqref{mainresult1}.  
In section 3 we apply this formula to calculate the butterfly velocities in M2-branes, M5 branes, Dp-branes and also string backgrounds. 
We find that the butterfly velocity in M2 equals to that in string background.  Also, the velocity in M5 equals to that in D4.   
Since the spacetime of string and D4 are those through double dimensional reduction from M2 and M5 backgrounds respectively, 
we conjecture that the butterfly velocity is invariant under a double-dimensional reduction. 
One more example involving the intersections F1$\bot$D4 and M2$\bot$M5, which are related also through a double-dimensional reduction, also supports our claim.  
In section 4, we develop a butterfly velocity formula of Gauss-Bonnet gravity with arbitrary matter fields. 
We present the detailed tensor analysis and obtain a compact and general formula given in \eqref{mainresult2}. 
For the special cases of planar, spherical or hyperbolic black holes, the formula becomes a simpler form described in \eqref{mainresult3}.  
In section 5 we first check that our formula reproduces the previous result in the literature without matter field [8,18]. 
We next  apply our formula to the Einstein-Gauss-Bonnet-Maxwell theory and Einstein-Gauss-Bonnet-scalar theory.  
Furthermore, we calculate the butterfly velocities in the Einstein-Gauss-Bonnet-Maxwell theory with spherical or hyperbolic black hole. 
We discuss future works in the  last section. 
In appendix A, we briefly describe the Kruskal coordinate in general geometry and derive some relations which are useful in computing butterfly velocity. 
In appendix B, we present some details in deriving our formula of the butterfly velocity in Einstain-Gauss-Bonnet gravity with arbitrary matter fields.

%%%%%%%%%%%%%%%%%%%%%%
%%%%%%%%%%%%%%%%%%%%%%

\section {Shock Wave Equation and Butterfly Velocity in Anisotropic Spacetime}
\subsection {Shock Wave Geometry and Shock Wave Equation} 
We will derive the formula of butterfly velocity in the following anisotropic background:
\be
ds^2&=&-a(r)f(r)dt^2+{dr^2\over b(r)f(r)}+\sum_{S=1}^n G^{(S)}_{ij}(r,x)\,dx_{(S)}^idx_{(S)}^j\\
G^{(S)}_{ij}(r,x)&=&h^{(S)}(r)g^{(S)}_{ij}(x)
\ee
where horizon locates at $r=r_H$. Note $f(r_H)=0$ while $a(r_H)\ne0,~b(r_H)\ne0$.  
In the original derivation [27],  Sfetsos considered an isotropic background representing the special case S=1. 
The authors in [17,22-26] considered the anisotropic case in flat space where $g^{(1)}_{ij}=g^{(2)}_{ij}=\delta_{ij}$.  
In our study, the relevant line element  (for example $g^{(2)}_{ij}(x)dx_{(2)}^idx_{(2)}^j$) will be able to describe  curved surface where  $g^{(2)}_{ij}\ne\delta_{ij}$. 
Note the associated temperature of above black hole or black brane is given by
\be
T={f\rq{}(r_H)\sqrt {a(r_H)b(r_H)}\over 4\pi}
\ee
Since the holographic geometry of the chaos covers two sides, we consider the line element expressed in the Kruskal coordinate : \footnote{In appendix A we describe the more properties of Kruskal coordinate. We also present several useful relations.}
\be
ds^2&=& -{4a(r)f(r)\over a(r_H)b(r_H)f\rq{}(r_H)^2}~e^{-f\rq{}(r_H)r_*\sqrt {a(r_H)b(r_H)}}dUdV+\sum_SG^{(S)}_{ij}(r,x)dx_{(S)}^idx_{(S)}^j\\
&=&2A(U,V)dUdV+\sum_Sh^{(S)}(UV)g^{(S)}_{ij}(x)dx_{(S)}^idx_{(S)}^j\\
UV&=&e^{{f\rq{}(r_H)}~r_*\sqrt {a(r_H)b(r_H)}},~~~U/V=e^{{f\rq{}(r_H)}~t\sqrt {a(r_H)b(r_H)}}
\ee
in which $r_*$ is the tortoise coordinate defined by $dr_*={dr\over f(r) \sqrt {a(r)b(r)}}$. 

To proceed we follow [5,10,11,27] to add a small null perturbation of asymptotic energy $E$.  
At later time the perturbation will follow null trajectories very close to the (past) horizon, where the trajectories become exponentially blue-shifted and the perturbation grow exponentially large, in contrast to what was expected in an earlier study [29]. 
After solving the associated Einstein's equation one obtain shock wave geometry shown in figure 1. 
\\
\\
\scalebox{0.4}{\hspace{12cm}\includegraphics{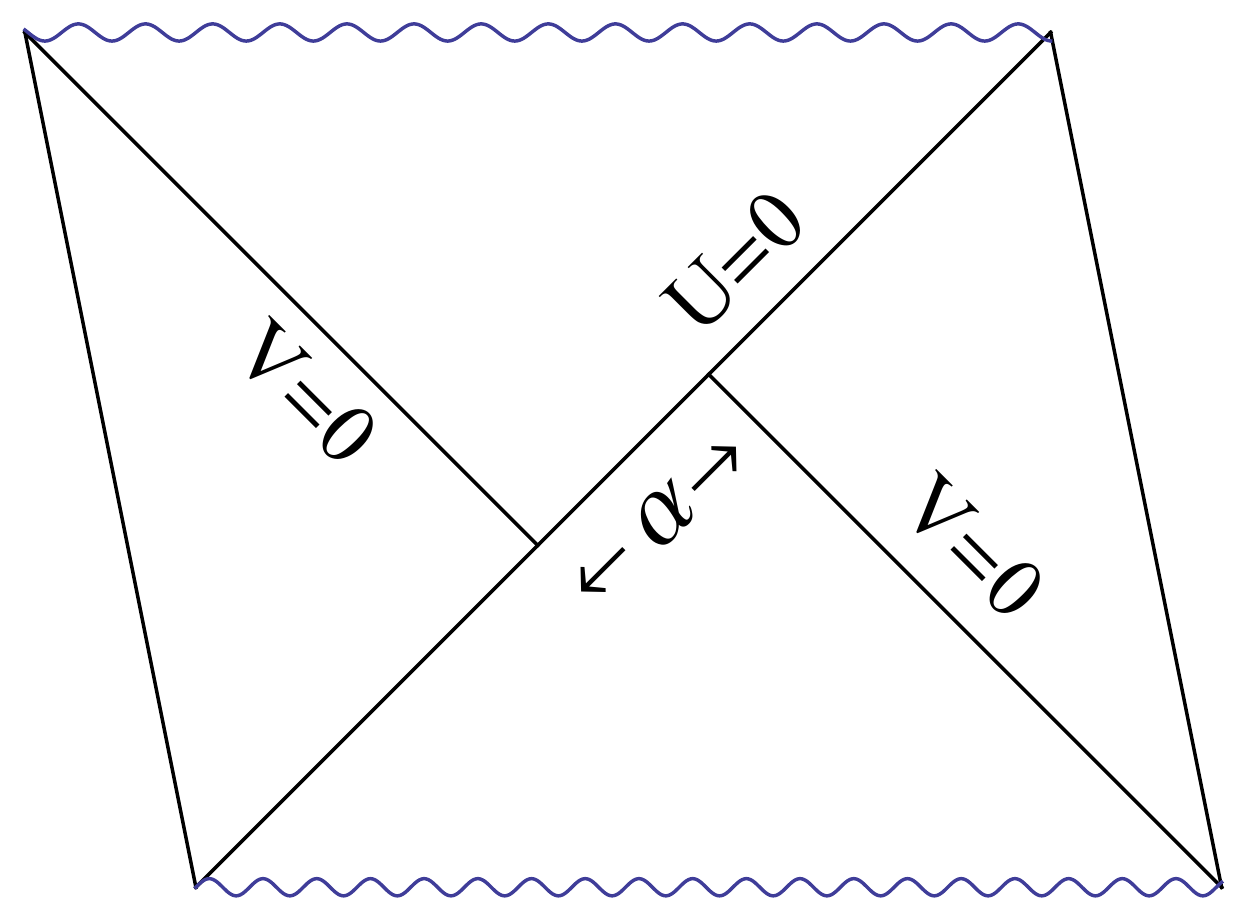}}
\\ 
{\it Figure 1:  Shock wave geometry : Penrose diagram of an eternal black hole perturbed by a shock wave.}
\\

We first describe the computation scheme of obtaining the shock wave equation in spacetime (2.1). 
This will enable us to find butterfly velocities in M-branes, D-branes and also string backgrounds.  
The scheme was clearly described in [27], and it had been applied to study the holographyic butterfly velocity with matter field in [10] recently.\\

The metric (2.1) describes the solution of Einstein equation with general form of stress tensor. We can express it as 
\be
G=T_{matter}&=&2T_{UV}(U,V,x)dUdV+T_{UU}(U,V,x)dUdU\nn\\
&&+T_{VV}(U,V,x)dVdV \nn\\
&&+\sum_ST^{(S)}_{ij}(U,V,x)dx_{(S)}^i dx_{(S)}^j
\ee where $G$ is the Einstein tensor.
Along the arguments of Dray and G. t'Hooft [29], for $U < 0$ the spacetime is described by (2.5).  
After adding a small null perturbation, for $U > 0$ the spacetime is still described by (2.5) but $V$ is shifted by 
\be
V \rightarrow V + \alpha(x)
\ee
as shown in figure 1 . 
The function $\alpha(x)$ will be determined by a shock wave equation which is to be determined.
The resulting metric and energy momentum tensor are
\be
ds^2&=&2A(U,V +\Theta \alpha(x))dU(dV +\Theta  \alpha'(x)dx)\nn\\
&&+\sum_SG^{(S)}_{ij}(U,+\Theta \alpha(x),x)dx_{(S)}^idx_{(S)}^j\\
T_{matter}&=&2T_{UV}(U,V+\Theta \alpha(x),x)dU(dV +\Theta  \alpha'(x)dx)\nn\\
&&+T_{UU}(U,V+\Theta \alpha(x),x)dUdU\nn\\
&&+ T_{VV}(U,V+\Theta \alpha(x),x)(dV +\Theta  \alpha'(x)dx)^2\nn\\
&&+ \sum_ST^{(S)}_{ij}(U,V+\Theta \alpha(x),x)dx_{(S)}^i dx_{(S)}^j~~~~~~~~
\ee
where $\Theta=\Theta(U)$ is a step function and 
\be
 \alpha'(x)dx=\sum_{i,S}{\partial \alpha(x)\over \partial x_{(S)}^i}~dx_{(S)}^i 
\ee
As described by Sfetsos [27], in terms of  the new coordinates
\be
\hat U=U,~~~\hat V=V+\Theta \alpha(x),~~~\hat x_{(S)}^i =x_{(S)}^i~~(i.e.~\hat X_\mu=X_\mu) 
\ee
the metric and stress tensor can be expressed by the following simple forms:
\be
\label{simpleform}
ds^2&=&2\hat A(\hat U,\hat V)d\hat Ud\hat V+\sum_S\hat G^{(S)}_{ij}(\hat U,\hat V,\hat x) d\hat  x_{(S)}^id\hat x_{(S)}^j - 2\hat A~\hat \alpha(\hat x)\hat \delta(\hat U) d\hat U^2\\
T_{matter}&=&2\Big(\hat T_{\hat U\hat V}-T_{\hat V\hat V}~\hat \alpha~\hat \delta\Big) d\hat Ud\hat V \nn\\
&&+\Big(T_{\hat U\hat U}+T_{\hat V\hat V}~\hat \alpha^2\hat \delta^2 -2T_{\hat U\hat V}~\hat \alpha ~\hat \delta\Big)d\hat U^2\nn\\
&&+ T_{\hat V\hat V}d\hat V^2 +\sum_S\hat T^{(S)}_{ij}d\hat x_{(S)}^i d\hat x_{(S)}^j
\ee
Following [8,27], we add an extra stress tensor 
\be
T_{(shock)\hat U\hat U}=E\,e^{2\pi t\over \beta}~a(x)~\delta(U)
\ee  to produce the shock wave geometry.

Now we have to solve the Einstein equation  (we will drop the hat notation in what follows.)
\be
G=T_{matter}+T_{shock}
\ee
The key point  is that using un-perturbed Einstein equation $G^{(0)}_{VV} = T^{(0)}_{VV}$ then  [21]
\be
T^{(0)}_{VV}=0
\ee
because $G^{(0)}_{VV}=0$ in the model spacetime.  
Therefore, $T_{matter}$ remains only one term that is linear in perturbation: $-2T_{\hat U\hat V}~\hat \alpha ~\hat \delta~d\hat U^2$ . 
The shock wave equation can be written as 
\be
\label{sweq}
G^{(1)}_{UU}+2G^{(0)}_{UV}~ \alpha(x) ~\delta(U)=CE\,e^{2\pi t\over \beta} ~a(x)~\delta(U)
\ee
where $G^{(1)}_{UU}$ is the first-order correction of the Einstein tensor from the metric  (2.13).

%%%%%%%%%%%%%%
%%%%%%%%%%%%%%

\subsection{Butterfly Velocity Formula in Anisotropic Spacetime}

According to the scheme described above, the main problem is to calculate the tensors $G^{(1)}_{UU}$  and $G^{(0)}_{UV}$.  
For the case of S=1, the above two tensors were presented in [27]. 
Here we are interested in a more general case.

We first need the following tensor properties in un-perturbed spacetime (2.5):
\be
G^{(0)}_{VV}(0)&=&0\\
G^{(0)}_{UV}&=&-{A'(0)\over A(0)}-\sum_S dim(S){h^{'(S)}(0)\over 2 h^{(S)}(0)}-{1\over 2}A(0)~R^{(0)}
\ee 
They be found through straightforward algebra calculations. 
Note $dim(S) = g^{(S)ij} g^{(S)}_{ij}$ arises from the tensor contraction from different species. 
We have considered values on the horizon, $U=0$.

In contrast to performing straightforward but tedious algebra calculations, we will calculate $G^{(1)}_{UU}$ starting from the following basic relation: 
\be
\delta R_{ab}&=&{1\over 2}\Big(\nabla^c\nabla_a\delta g_{cb}+\nabla^c\nabla_b\delta g_{ca}-g^{cd}\nabla_a\nabla_b\delta g_{cd}-\nabla^2\delta g_{ab}\Big)
\ee
Now from $\delta g_{ab}=-2A~\alpha(x)\delta(U) \delta_{aU}\delta_{bU}$ and $g^{(0)}_{UU}=g^{(0)UU}=0$ we find 
\be
G^{(1)}_{UU}&=&\nabla^\lambda\nabla_U \delta g_{\lambda U}-{1\over 2}\nabla^2 K-{1\over 2}K R^{(0)} \\
K&=&-2A(UV)~\alpha(x)\delta(U)
\ee
Note that in the above relations the covariant derivative acts on un-perturbed background. 
It is interesting that the term ${1\over 2}K R^{(0)}$ is canceled by ${1\over 2}A(0)~R^{(0)}$ in (2.20), according to the shock wave equation (2.18). 
To proceed we find
\be
&&\nabla^2 K={1\over A~\prod_{ s}\sqrt {h^{(S)} g^{( S)}_{ij}}} \partial_a ~\Big[\Big(A~\prod_{ S}\sqrt {h^{(S)} g^{( S)}_{ij}}\Big)g^{ab} \partial_a K\Big]\\
&&~~~=\Big({4A'(0)\over A(0)}+\sum_S {dim(S)h^{'(S)}(0)\over  h^{(S)}(0)}\Big) \alpha(x)\delta(U)-2A(0)\delta(U)\sum_S {\Delta^{(S)}\alpha(x)\over h^{(S)}}~~~~\\
&&\nabla^\lambda\nabla_U\delta g_{\lambda U}=\Big({2A'(0)\over A(0)}+\sum_S dim(S){h^{'(S)}(0)\over  h^{(S)}(0)}\Big)~\alpha(x)\delta(U)
\ee
where $g^{ab}$ is the metric given in (2.5). The Laplacian is
\be
\Delta^{(S)}~\alpha(x)={1\over \sqrt {g^{(S)}}} ~\partial_j^{(S)} \Big(\sqrt {g^{(S)}}~ {g^{(S)ij}}~ \partial^{(S)}_j  \alpha(x)\Big)
\ee
To obtain above result we have used $\delta'(U)=-\delta(U)/U$.

Collecting all the above results, we finally obtain
\be
&&G^{(1)}_{UU}+2G^{(0)}_{UV}~\alpha(x)\delta(U)\nn\\
&=&\delta(U)\sum_S {A(0)\over h^{(S)}(0)}~\Delta^{(S)}~\alpha(x) -\sum_S dim(S){h^{'(S)}(0)\over  2h^{(S)}(0)}\alpha(x)\delta(U)
\ee
Note $R^{(0)}$ parts are canceled out. 
The shift function $\alpha(t,x)$ is determined by the perturbation function $a(x)$ through the shock wave equation
\be
\Big[A(U_H)\sum_S(h^{(S)})^{-1}\Delta^{(S)}-{1\over2}\sum_SG^{(S)ij}(U_H,x)G^{'(S)}_{ij}(U_H,x)\Big]~\alpha(t,x)= E\,e^{2\pi t/\beta}~a(x)~
\ee
where $G^{'(S)}_{ij}(U_H,x)={\partial G^{(S)}_{ij}(u,x)\over \partial U}_{|U=U_H}$.

To obtain a simple formula of butterfly velocity we can consider the  case in which the local source is $a(x)=\delta(x_i^{(Q)})$. 
Using the relation (2.2) we find
\be
\label{mainresult1}
\Big[\Delta^{(Q)}-h^{(Q)}\sum_S dim(S)~{h^{'(S)}\over 2A~ h^{(S)}}\Big]_{U_H} ~\alpha(t,x_i^{(Q)}) = E\,e^{2\pi t/\beta}~{h^{(Q)}(U_H)\over A(U_H)}~\delta(x_i^{(Q)})
\ee
where $dim(S)$ is the spatial dimension of $x_{(S)}^i$.  
The velocity along the direct $x_{(Q)}^i$ in the anisotropic spacetime (2.1) can be found to be
\be
v_B^{(Q)}&=&{2\pi kT\over M_{(Q)}}\\
M^2_{(Q)}&=&h^{(Q)}(r_H)\sum_S dim(S)~{b(r_H) f'(r_H)h^{'(S)}(r_H)\over 4 h^{(S)}(r_H)}
\ee
where we have used 
\be
h'(U=0)=r_Hh'(r_H),~~~A(U=0)={2r_H\over b(r_H) f'(r_H)}
\ee
which are derived in appendix A. 
Note that our formula (2.30) reduces to [17,22-26] and [27] in the flat anisotropic space and curved isotropic space, respectively. 
In the next section we will use our new formulae to evaluate the butterfly velocities in the brane geometry. 

%%%%%%%%%%%%
%%%%%%%%%%%%

\section{Butterfly Velocity in Brane Geometry}
\subsection{M5 and D4 Backgrounds}
In this section we use our general butterfly velocity formula to calculate the butterfly velocities in M-branes, D-branes and string backgrounds. 
The $N_5$ black M5-branes solution is given by [30,31]
\be
ds^2_{M5}&=&H^{-1\over3}\left(-f(r)dt^2 +dx_1^2 +dx_2^2+dx_3^2 +dx_4^2 +dx_5^2\right) + H^{2\over3} \left({dr^2\over f(r)}+r^2 d\Omega_4^2\right)~~
\ee
where $H$ is the harmonic function defined by 
\be 
H = 1+ {N_5\over r^3}
\ee
The function $f(r)$ specified by the horizon at $r_H$ is 
\be
f(r)=1- {r_H^3\over r^3}
\ee
We can approximate $H\sim {N_5\over r^3}$ in the ``near-horizon" limit. 
On the other hand, the spacetime of a stack of  $N_4$ black D4-branes  (in the Einstein frame) is given by
\be
ds_{D4}^2&=&-H^{-3\over8}f(r) dt^2 + H^{-3\over8}\left(dx_1^2+dx_2^2+dx_3^2 +dx_4^2\right)+H^{5\over8} \left({dr^2\over f(r)}+r^2 d\Omega_4^2\right)\\
H &=& 1+ {N_4\over r^3},~~~~~f(r)=1- {r_H^3\over r^3}
\ee
In the ``near-horizon" limit, $H\sim {N_4\over r^3}$. 
Using these metrices, we calculate the black-brane temperature, $T(r_H)$, the butterfly velocities $v_B$ along $(x_1\cdot\cdot\cdot x_5)$ and along $\Omega_4$ in M5. We can also compute the velocities along $(x_1\cdot\cdot\cdot x_4)$ and along  $\Omega_4$ in D4. 
We collect the relevant functions in the table 1.
\\

{\it Table 1:  Parameter functions for M5 and D4.}
\\
\scalebox{0.7}{\hspace{0cm}\includegraphics{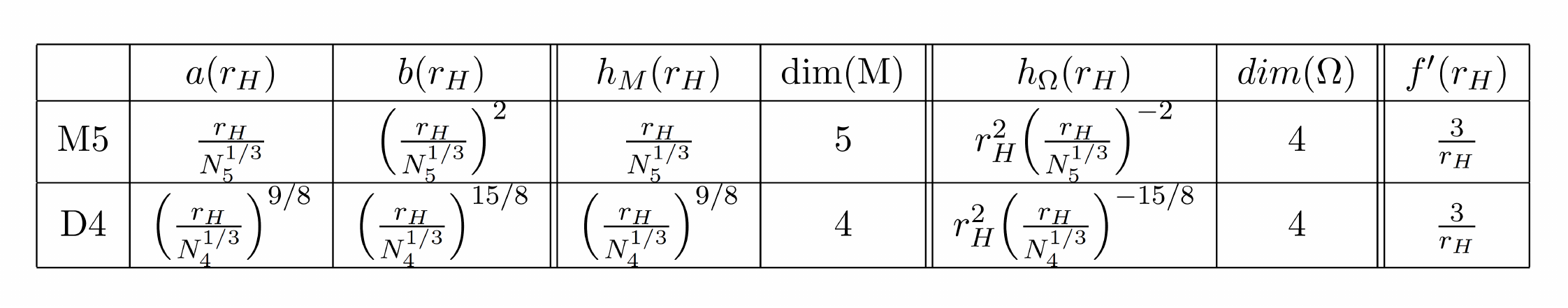}}
\\          
It seems striking that although the parameter functions for M5 and D4 are quite different, the resulting butterfly velocities and temperature, given below,
\be
\ba {cccc}
&~~T(r_H)~~&~~v_B(M)~~&~~v_B(\Omega)\\
M5&{3\over4\pi} \sqrt{r_H\over N_5}&\sqrt{3\over 5}&{4\pi \over \sqrt{15}}T\\
D4&{3\over4\pi} \sqrt{r_H\over N_4}&\sqrt{3\over 5}&{4\pi \over \sqrt{15}}T\\
\ea
\ee
are the same.   Let us make following comments :

1. For temperature to be the same, one needs the same number of branes for M5 and D4. The butterfly velocities are the same independent of N.

2. While the butterfly velocity $v_B(M)$ satisfies upper bound  $v_B(\Omega)$ violates  it. This violation does not contract to [4] since that paper assumes space  isotropy [15,17].

3. Notice that through a double-dimensional reduction M5 becomes D4.  

4. For completeness we present in below the butterfly velocity in Dp brans background
\be
\ba {cccc}
&~~T(r_H)~~&~~v_B(M)~~&~~v_B(\Omega)\\
Dp&{7-p\over4\pi \sqrt{Np}}~r_H^{5-p\over 2}&\sqrt{7-p\over 9-p}&{4\pi T\over \sqrt{(7-p)(9-p)}}
\ea
\ee
%%%%%%%%%%%%%%
%%%%%%%%%%%%%%
\subsection{M2 and Fundamental String (F1) Backgrounds}
Now we consider the systems of M2 and fundamental string (F1) backgrounds. 
The relevant functions are collected in the table 2.
\\

{\it Table 2:  Parameter functions for M2 and F1.}
\\
\scalebox{0.7}{\hspace{0cm}\includegraphics{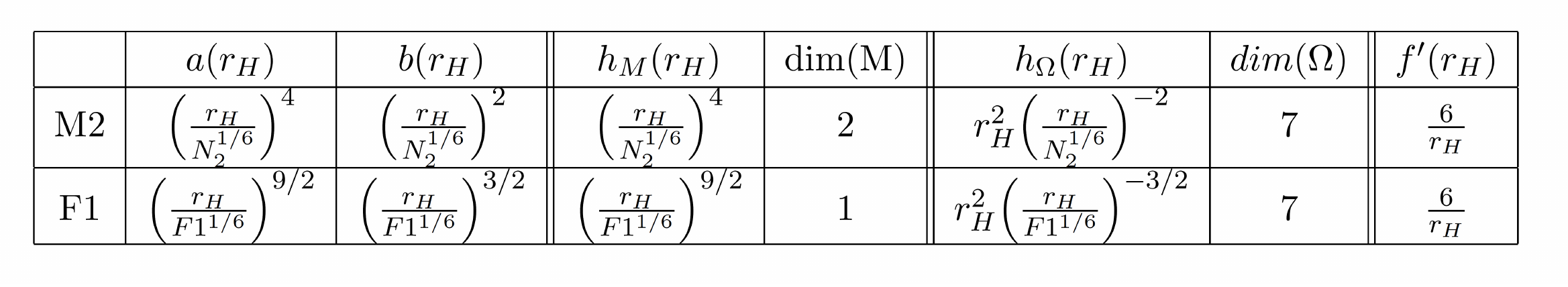}}
\\          
We again observe that although the parameter functions for M2 and F1 are different, the temperature and butterfly velocities, shown in below,
\be
\ba {cccc}
&~~T(r_H)~~&~~v_B(M)~~&~~v_B(\Omega)\\
M2&{3r_H^2\over2\pi \sqrt {N_2}}&{\sqrt3\over 2}&{\pi \over \sqrt3}T\\
F1&{3r_H\over2\pi \sqrt{F_1}}&{\sqrt3\over 2}&{\pi \over \sqrt3}T\\
\ea
\ee
have the same behaviour.
Since that through a double-dimensional reduction M2 (M5) becomes F1 (D4) one tempts to conjecture that the butterfly velocity is generally invariant under a double dimensional reduction. Let us consider one more example in the next section.

%%%%%%%%%%%%%%
%%%%%%%%%%%%%%
%%%%%%%%%%%%%%
\subsection{Intersections M2$\bot$M5 and F1$\bot$D4 Backgrounds}

We consider the brane intersection M2$\bot$M5 system where M2-branes lie in ($w,x$) and M5-branes locate at ($w,y_1...y_4$) i.e.,
\be
\ba {cccccccccccc}
&t&w&x&y_1&y_2&y_3&y_4&z_1&z_2&z_3&z_4\\
M2&\bullet&\bullet&\bullet&&&&&&&&\\
M5&\bullet&\bullet&&\bullet&\bullet&\bullet&\bullet&&&&\\
\ea
\ee 
Upon a double-dimensional reduction, the space $w$ is wrapped, and the geometry becomes the intersection F1$\bot$D4, i.e.,
\be
\ba {ccccccccccc}
&t&x&y_1&y_2&y_3&y_4&z_1&z_2&z_3&z_4\\
F1&\bullet&\bullet&&&&&&&&\\
D4&\bullet&&\bullet&\bullet&\bullet&\bullet&&&&\\
\ea
\ee 
The metric of black intersection M2$\bot$M5 is [30,31]
\be
ds^2_{M2\bot M5}&=&H_2^{-2\over3}H_5^{-1\over3}[(-f(r)dt^2 +dw^2] +H_2^{-2\over3}H_5^{2\over3} dx^2+H_2^{1\over3}H_5^{-1\over3}d\vec y^2\nn\\
&&+H_2^{1\over3}H_5^{2\over3}[f^{-1}(r)dr^2+r^2d\Omega_3^2]
\ee
where
\be
H_2&=&1+{{N_2}\over r^2},~~H_5=1+{{N_5}\over r^2},~~~f(r)=1-{{r_H}^2\over r^2}
\ee
In the Einstein frame, the metric of the intersection F1$\bot$D4 is 
\be
ds^2_{F1\bot D4}&=&-H_1^{-3\over4}H_4^{-3\over8}f(r)dt^2  +H_1^{-3\over4}H_4^{5\over8} dx^2+H_1^{1\over4}H_4^{-3\over8}d\vec y^2\nn\\
&&+H_1^{1\over4}H_4^{5\over8}[f^{-1}(r)dr^2+r^2d\Omega_3^2]
\ee
where
\be
H_1&=&1+{{F_1}\over r^2},~~H_4=1+{{N_4}\over r^2},~~~f(r)=1-{{r_H}^2\over r^2}
\ee In the ``near horizon" limit, the relevant functions are collected in the table 3.
\\

{\it Table  3:  Parameter functions for M2$\bot$M5  and F1$\bot$D4.}
\\
\scalebox{0.7}{\hspace{0cm}\includegraphics{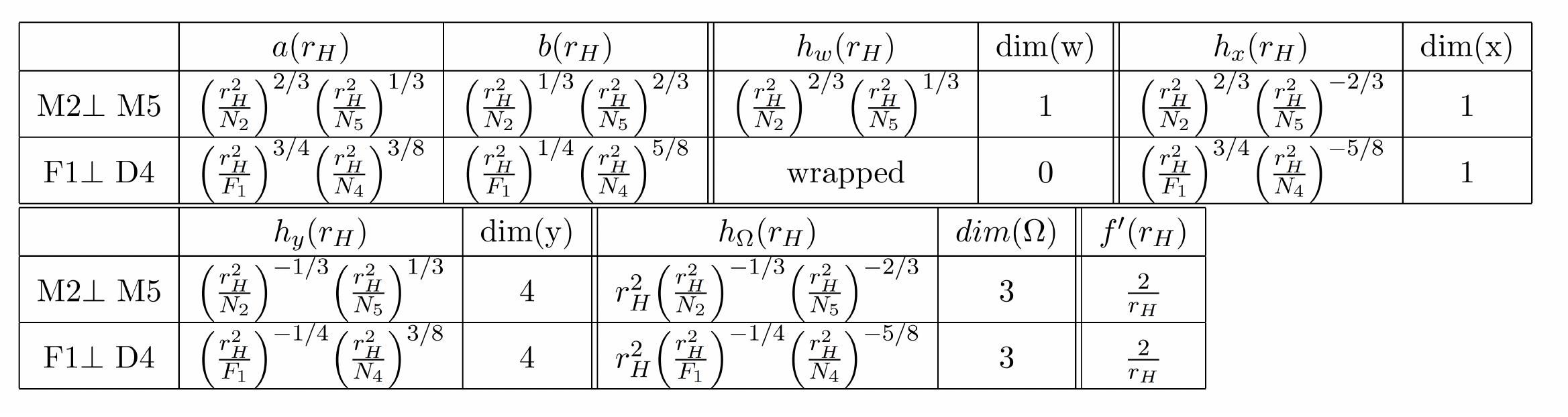}}
\\          
Although the parameter functions for M2$\bot$M5  and F1$\bot$D4 are different, the temperature and butterfly velocities
\be
\ba {ccccc}
&~~T(r_H)~~&~~v_B(x)~~&~~v_B(y)~~&~~v_B(\Omega)\\
 M2\bot M5&{r_H\over2\pi \sqrt{N_2N_5}}&2\pi T\sqrt{N_2}&2\pi T\sqrt{N_5}&2\pi T\\
F1 \bot D4&{r_H\over2\pi \sqrt{F_1N_4}}&2\pi T\sqrt{F_1}&2\pi T\sqrt{N_4}&2\pi T
\ea
\ee
again have same behaviors.  Thus, the butterfly velocity is again invariant under a double-dimensional reduction.

%%%%%%%%%%%%
%%%%%%%%%%%%
\section{Formula of Butterfly velocity in  Einstein-Gauss-Bonnet Gravity with Matter Fields}
\subsection{Formula in Arbitrary Spacetime}
The Lagrangian we consider will contain a curvature scalar, a cosmological constant, a Gauss-Bonnet term and matter fields:
\be
{\cal L}&=&R+{d(d+1)\over \ell_{AdS}^2}+{{\gamma_{GB}}\over 2}~\Big(R_{abcd}R^{abcd} - 4R_{ab}R^{ab}+R^2\Big)+{\cal L}_{matter}
\ee
where $\ell_{AdS}$ is the AdS radius.  
The associated gravity equation is given by
\be
G_{ab}+{2d(d+1)\over  \ell_{AdS}^2}g_{ab}-\gamma_{ GB}\Big(H_{ab}-{1\over 4}H g_{ab}\Big)=T_{ab}
\ee
where we denote
\be
H_{ab}=R_{acde}R_b^{~cde}-2R_{acbd}R^{cd}-2R_{ac}R_b^{c}+R_{ab}R
\ee
The Gauss-Bonnet term is a special combination of curvatures in which its field question contains only second-order derivatives. All higher-derivatives terms are cancelled out.  
In four-dimensions, the Gauss-Bonnet term is a topological invariant so it does not enter dynamics. We will consider the theory in d+2 dimensions  \footnote{Note that the total dimension of spacetime D in [8] and [18] is D=1+d, while in here we let D=2+d. 
In our notation the Laplacian is then denoted as  $\Delta^{(d)}$ which is more convenient in our discussion.}.  
We assume that the solution expressed in Kruskal coordinates could be written as
\be
\label{unpurtubed}
ds^2=2A(U,V)dUdV+h(U,V)\sum_{i,j=1}^d g_{ij}(x)dx^idx^j
\ee
Above metric lead to 
\be
H_{VV}^{(0)}-{1\over 4}H^{(0)} g^{(0)}_{VV}=0
\ee
and  condition $T_{VV}^{(0)}=0$ is automatically satisfied.  
After perturbing the metric, we have an extra contribution on the $g^{(1)}_{UU}$ conponent, which is linear in $\alpha$ as described in \eqref{simpleform}. 

In Gauss-Bonnet gravity, according to the scheme described in section 2, the shock-wave equation  \label{sweq} becomes
\be
\Big[H_{UU}^{(1)}-{1\over 4}H^{(0)} g^{(1)}_{UU}\Big] +2\Big[H_{UV}^{(0)} -{1\over 4}g^{(0)}_{UV}H^{(0)}\Big]\alpha(x)\delta(U)=E^{2\pi t/\beta}~a(x)~\delta(U)
\ee
Using  the definitions  $ g^{(1)}_{UU}=-2A\alpha(x)\delta(U)$ and $g^{(0)}_{UV}=A$,  the second terms in the first bracket and second bracket are canceled exactly. 
This property is the same as what happens in Einstein gravity where the terms ${1\over2}g^{(0)}_{UV} R^{(0)}$ are canceled out  in section 2. 
The shock wave equation for Gauss-Bonnet gravity now has a very simple form:
\be
H_{UU}^{(1)} + 2H_{UV}^{(0)}~\alpha(x)\delta(U)=E^{2\pi t/\beta}~a(x)~\delta(U)
\ee
Now we only have to calculate $H_{UU}^{(1)}$ and $H_{UV}^{(0)}$, just like as we only had to calculate $G_{UU}^{(1)}$ and $G_{UV}^{(0)}$ in \eqref{simpleform} for the  Einstein gravity theory. 

To proceed, we find the following general relations:
\be
H^{(0)}_{UV}&=&- R^{(0)}\Big[{A'\over A}+{(d-2)h'\over 2h}\Big]-{2A^{'2}\over A^3}-{(3d-2)A'h'\over A^2h}-{d(d-3)h'^{2}\over 2Ah^2}\\
H^{(1)}_{UU}&=&(R_{Ucde}R_U^{~cde})^{(1)}- 2(R_{UcUd} R^{cd})^{(1)} -2(R_{Uc}R_U^{c})^{(1)} +(R_{UU}R)^{(1)}
\ee
where, up to a factor $\delta(U)$
\be
(R_{Ucde}R_U^{~cde})^{(1)}&=&-{4A^{'2}\over A^3}\alpha-{2h'\over h^2}\Delta^{(d)}\alpha\\
\label{old411}
(R_{UcUd} R^{cd})^{(1)}&=&AR^{(0)ij} \nabla^{(d)}_i\nabla^{(d)}_j\alpha(x) +{A'\over Ah}\Delta^{(d)}\alpha(x)\nn\\
&&+\Big({R^{(0)}h'\over 2h}-{2A'^2\over A^3}+{dh{'^2}\over  Ah^2}-{(3d-2)A'h'\over 2A^2h}\Big)\alpha(x)\\
(R_{Uc}R_U^{c})^{(1)}&=&-\Big({2A^{'2}\over A^3}+{dA'h'\over A^2h}\Big)\alpha-\Big({2A^{'}\over Ah}+{dh'\over h^2}\Big)\Delta^{(d)}\alpha\\
(R_{UU}R)^{(1)}&=&\Big({2A^{'}\over A}+{dh'\over 2h}\Big)R^{(0)}\alpha+{A\over h}R^{(0)}\Delta^{(d)}\alpha
\ee
The covariant derivative $\nabla^{(d)}_i$ and Laplacian $\Delta^{(d)}$ are  defined in the  d-dimensional metric, $ds^2=g_{ij}dx^idx^j$.  
$R^{(0)}$ is the 2+d dimensional Ricci scalar calculated by un-perturbative metric \eqref{unpurtubed}. 
Note that all values are evaluated at the horizon. 
It will require further analysis to derive the relation $(R_{UcUd} R^{cd})^{(1)}$, which we present them in the appendix B.  It is interesting to see that in the Einstein gravity the term  containing derivatives of $\alpha(x)$ are formed as the Laplacian, $\Delta^{(d)} \alpha(x)$,  while in the Gauss-Bonnet gravity it  appears  a new form $R^{ij}\nabla^{(d)}_i\nabla^{(d)}_j\alpha(x)$. 

Collect above calculations, the shock wave equation of Einstein-Gauss-Bonnet gravity  from
\be
\Big[G^{(1)}_{UU}+2G^{(0)}_{UV} \alpha(x) \delta(U)\Big]-\gamma_{ GB}\Big[H_{UU}^{(1)} + 2H_{UV}^{(0)}~\alpha(x)\delta(U)\Big]=E\,e^{2\pi t/\beta}~a(x)~\delta(U)
\ee
reduces to the following  formula:
\be
&&\Big[\Delta^{(d)} \alpha(x)-{dh'\over 2A}\alpha(x)\Big]_{U=0}\nn\\
&&-\gamma_{ GB}\Big[ -2hR^{(0)ij} \nabla^{(d)}_i\nabla^{(d)}_j\alpha(x) +\Big({R^{(0)}}+{2A'\over A^2}+{2(d-1)h'\over  Ah}\Big)\Delta^{(d)}\alpha(x)\nn\\
&&-\Big({R^{(0)}(d-2)h'\over 2A}+{(d-2)A'h'\over A^3}+{d(d-2)h{'^2}\over  A^2h}\Big)\alpha(x)\Big]_{U=0}= {h(0)\over A(0)}E\,e^{{2\pi t\over \beta}}~a(x)\nn\\
\ee
evaluated in  Kruskal coordinate i.e.,  $A=A(UV)$ and $h=h(UV)$.  Using the tensor relations
\be
R^{(0)ij} \nabla^{(d)}_i\nabla^{(d)}_j\alpha(x)&=&{1\over h^2}R^{(d)ij} \nabla^{(d)}_i\nabla^{(d)}_j\alpha(x)-{h'\over  Ah^2}\Delta^{(d)}\alpha(x)\\
R^{(0)}&=&{R^{(d)}\over h}-{2A'\over A^2}-{2dh'\over  Ah}
\ee
the butterfly velocity formula has a simple form  
\be
\label{mainresult2}
&&\Big[\Delta^{(d)} \alpha(x)-{dh'\over 2A}\alpha(x)\Big]_{U=0}\nn\\
&&-\gamma_{ GB}\Big[ -{2R^{(d)ij}\over h} \nabla^{(d)}_i\nabla^{(d)}_j\alpha(x)
+{R^{(d)}\over h}\Delta^{(d)} \alpha(x)-{(d-2)h'R^{(d)}\over 2Ah}\alpha(x)\Big]_{U=0}= {h(0)\over A(0)}E\,e^{{2\pi t\over \beta}}~a(x)\nn\\
\ee
in which $R^{(d)ij}$ and $R^{(d)}$ are  defined in the  d-dimensional metric  $ds^2=g_{ij}dx^idx^j$, 
%%%%%%%%%%%%%%%%
%%%%%%%%%%%%%%%%
\subsection{Formula in  Planar, Spherical and Hyperbolic Black Holes}
To further simplify \eqref{mainresult2}, let us consider 2+d dimensional planar, spherical or hyperbolic black holes.  The general metric is 
\be
g_{ij}(x)dx^idx^j&=&\left\{
 \ba {cc}
d\theta_1^2+d\theta_2^2+ \cdot\cdot\cdot+d\theta_d^2,&k=0\nn\\
 d\theta_1^2+\sin^2\theta_1(d\theta_2^2+\sin^2\theta_2(d\theta_3^2+ \cdot\cdot\cdot+\sin^2\theta_{d-1}d\theta_d^2),&k=1\nn\\
 d\theta_1^2+\sinh^2\theta_1(d\theta_2^2+\sin^2\theta_2(d\theta_3^2+ \cdot\cdot\cdot+\sin^2\theta_{d-1}d\theta_d^2),&~~k=-1\nn\\
 \ea
 \right.
 \\
\ee
then
\be
R^{(d)ij} \nabla^{(d)}_i\nabla^{(d)}_j\alpha(x)&=&k(d-1)\Delta^{(d)}\alpha(x)\\
R^{(d)}&=&kd(d-1)
\ee
Substituting these relations into \eqref{mainresult2} and with a help of appendix A  we obtain an amazingly simple expression for the shock-wave equation of Einstein-Gauss-Bonnet gravity with arbitrary matters:  
\be
\label{mainresult3}
\Big(1-k{(d-2)(d-1)\gamma_{GB}\over h(r_H)}\Big)\Big(\Delta^{(d)} -M^2\Big)\alpha(x)\sim \delta (x)
\ee
where
\be
M^2&=&{d~h'(0)\over 2A(0)}=d\pi h'(r_H)~T \\
\label{vB}
v_B&=&{2\pi T\over M}=\sqrt{4\pi T\over dh'(r_H)}
\ee
 
Let us summarize and make some remarks.

1. Shock wave equation \eqref{mainresult2} can be directly applied to Einstein-Gauss-Bonnet theory with arbitrary matters once the black hole metric being expressed as \eqref{unpurtubed}.  For planar, spherical or hyperbolic black holes we have a simple formula \eqref{mainresult3}.
  
2. The case of $k=0$  was studied in [8].

3.  The time for the function $\alpha(t,x)$  becomes order-one  after the perturbation is defined as the  ''screaming time''. Use the function in (1.1) we see that $\alpha(t,x)\sim e^{\lambda(t-t*-{|x|\over v_B})}$.  The formula \eqref{mainresult3} shows an overall factor which may be  regarded as rescales the perturbation source or effectively be absorbed  into the exponential in (1.1).  This then changes  the screaming time by the amount $\log\Big(1-k{(d-2)(d-1) \gamma_{GB}\over r_H^2}\Big)$  as first found in [8]. 
 
4. The factor $(d-2)(d-1)\gamma_{GB}$  vanishes for $d=1, 2$.  The reason is that :  (a) The case of $d=1$ is in fact planar and it shall be zero.  (b) The Gauss-Bonnet is topology if $d=2$, and zero value at $d=2$ reveals this property.

 In the next section we consider some examples to illustrate our formulae. 
%%%%%%%%%%%

\section{Butterfly Velocity in Einstein-Gauss-Bonnet Gravity with Matters}
\subsection{Planar Black Hole}
In this subsection we  consider the case in which the butterfly is propagating in the planar black hole background
\subsubsection{Einstein-Gauss-Bonnet Gravity}
Consider first the simplest theory without any matter. 
The planar black hole solution is [32-34]
\be
ds^2&=&-f(r)N_{\sharp}^2dt^2+f(r)^{-1}dr^2+r^2\sum_{i=1}^ddx^idx^i
\ee
with
\be
f(r)&=&{r^2\over 2\lambda}\Big[1-\sqrt{1-4\lambda(1-(r/r_H)^{-d-1})}\Big]\\
T&=&N_{\sharp}^2{(d+1)r_H\over 4\pi}\\
N_{\sharp}^2&=&{1\over 2}\Big(1+\sqrt{1-4\lambda}\Big),~~~\lambda\equiv\gamma_{GB}(d-1)(d-2)
\ee
We let the AdS radius $\ell_{AdS}=1$. The horizon locates on $r_H$.   
As discussed in [8], the presence of the parameter $N_{\sharp}$ in the metric rescales time so the temperature is $\beta=4\pi f'(1)/N_{\sharp}$.  
This rescaling implies that the butterfly velocities in Einstein gravity and Einstein-Gauss-Bonnet gravity without matters are
\be
v_B(0)&=&{\sqrt{(d+1)\over  2d}}\\
\label{vq1}
v_B(\gamma_{ GB}) &=&N_{\sharp}v_B(0)=\Big(1+ \sqrt{1-4(d-1)(d-2)\gamma_{ GB}} \Big)^{1/2}~v_B(0)
\ee
respectively. These results were first found in [8]. 
%%%%%%%%%%%%%%%%%%%%%%
%%%%%%%%%%%%%%%%%%%%%%
%%%%%%%%%%%%%%%%%%%%%%
\subsubsection{Einstein-Gauss-Bonnet-Maxwell Gravity}
For the  Einstein-Gauss-Bonnet Gravity theory with Maxwell field, one adds a matter Lagrangian 
\be
{\cal L}_{matter}=-{1\over 4g^2}F_{\mu\nu}F^{\mu\nu}
\ee
The associated planar black hole solution then is [32-34]
\be
ds^2&=&-f_q(r)N_{\sharp}^2dt^2+f_q(r)^{-1}dr^2+r^2\sum_{i=1}^ddx^idx^i\\
A_t&=&\mu\Big(1-(r/r_H)^{-(d-1)}\Big)
\ee
where 
\be
f_q(r)&=&{r^2\over 2\lambda}\Big[1-\sqrt{1-4\lambda(1-(1+q)(r/r_H)^{-d-1} + q(r/r_H)^{-2d})}\Big]\\
T&=&N_{\sharp}^2{\Big((d+1)+(1-d)q\Big)r_H\over 4\pi}\\
q&=&{d-1\over d}{\mu^2\over g^2r_H^2}
\ee
The parameters $q$ and $\mu$ denote the strength of Maxwell fields. 
The values of $N_{\sharp}$ and $\lambda$ are defined before.  

We find the butterfly velocities in the Einstein-Maxwell Gravity theory and Einstein-Gauss-Bonnet-Maxwell Gravity theory are  given by
\be
v^{(q)}_B(0)&=&{\sqrt{(d+1)+(1-d)q\over  2d}}\\
\label{vq2}
v^{(q)}_B(\gamma_{ GB}) &=&N_{\sharp}v^{(q)}_B(0)=\Big(1+ \sqrt{1-4(d-1)(d-2)\gamma_{ GB}} \Big)^{1/2}~v^{(q)}_B(0)
\ee
respectively.  
When $q=0$ above results reduce to previous case. Comparing \eqref{vq1} and \eqref{vq2} we observe that the ratio of butterfly velocities between that with and without Gauss-Bonnet term is in fact ``universal", in the sense that it does not depend on the strength of Maxwell field. 

%%%%%%%%%%%%%%%%%%%%%%
%%%%%%%%%%%%%%%%%%%%%%
\subsubsection{Einstein-Gauss-Bonnet-Scalar Gravity}
Consider the  Einstein-Gauss-Bonnet Gravity theory with scalar fields.  The matter field Lagrangian is
\be
{\cal L}_{matter}=-{1\over 2}\partial_\mu\Phi \partial^\mu \Phi+{\xi\over 2}R\Phi+U(\Phi)
\ee
Choosing the particular values of  parameters
\be
\gamma_{GM}&=&{1\over 2(d-1)(d-2)},~~~~\Lambda=-{d(d+1)\over 4}\\
\xi&=&{d\over 4(d_2)},~~~~~~~~~~~~~U(\Phi)={d^2\over 32}\Phi^2
\ee
where $\Lambda$ is the cosmological constant,  the exact black hole solution can be described by [35,36]
\be
ds^2&=&-f(r)dt^2+f^{-1}dr^2+r^2\sum_{i=1}^ddx^idx^i\\
f(r)&=&r^2(1-(r/r_H)^{-d\over 2}),~~~~T={r_H~d\over 8\pi }
\ee 
We find the butterfly velocity is
\be
v_B={1\over 2}
\ee
Note that in this model the coupling of Gauss-Bonnet term $\gamma_{GB}$ is fixed in (5.16).

\subsection{Spherical and Hyperbolic Black Holes}

Next we consider non-planar black hole backgrounds, including spherical and hyperbolic geometry.

\subsubsection{Einstein Gravity}

Consider first the simplest case.  
The metric of a  black hole in Einstein Gravity theory  is
\be
ds^2&=&-f(r)dt^2+f_q(r)^{-1}dr^2+r^2\sum_{i,j=1}^dg_{ij}dx^idx^j\\
f(r)&=&k+r^2-\Big({r\over r_H}\Big)^{1-d}(k+r_H^2)\\
T&=&{(d+1)r_H^2+(d-1)k\over 4\pi r_H}
\ee
where
$k=0,1,-1$ describe flat, spherical and hyperbolic black hole, respectively.  
The  relation between horizon radius and the black hole temperature is more complicated and is given by
\be
r_H={2\pi T+\sqrt{4\pi^2 T^2+k(1-d^2)}\over (1+d)}
\ee
The butterfly velocity becomes
\be
v_B(k)&=&\sqrt{{2\pi (1+d)T\over2d\pi T+d\sqrt{k(1-d^2)+4\pi^2T^2}}}
\ee
which implies that
\be
v_B(k=0)&=&\sqrt{ (1+d)\over2d}\\
v_B(k)&=&\sqrt{(1+d)\over2d}+k {(d-1)(d+1)^{3\over2}\over 32  \pi^2\sqrt{2d}~ T^2}+{\cal O}(T^{-3})
\ee
where the first relation was found earlier in [8].
%%%%%%%%%%%%%%%%%%%%%%
%%%%%%%%%%%%%%%%%%%%%%

\subsubsection{Einstein-Gauss-Bonnet-Maxwell Gravity}

Finally, we consider  2+d dimensional planar, spherical or hyperbolic black hole solutions in Einstein-Gauss-Bonnet-Maxwell gravity. 
After taking a proper limit, i.e. $k=0$ and/or $q=0$,  the result obtained in what follows  reproduces all previous results, as they must be.

The metric is 
\be
ds^2&=&-f(r)N_\sharp^2dt^2+f^{-1}dr^2+r^2 d\Omega^2_d\\
f(r)&=&k+{r^2\over 2\lambda}\left[1-\sqrt{1-4\lambda\Big(1+({r\over r_H})^{-2d}\left(q-{({r\over r_H})^{-1+d}\Big((1+q)r_H^4+k r_H^2+k^2\lambda\Big)\over r_H^4}\right)}~\right]\nn\\
\label{T}
T&=&N_\sharp{((d+1)+(1-d)q)r_H^4+(d-1)kr_H^2+(d-3)k^2\lambda \over 4r_H \pi(r_H^2+2k\lambda) }
\ee
To find the butterfly velocity we solve \eqref{T} to express the horizon radius $r_H$ in terms of the black hole  temperature $T$, and then substitute it into \eqref{vB}. Since the exact expression is complicated we consider the high-temperature expansion. We find
\be
r_H={4\pi T\over  (1+d)+(1-d)q}+ {k[1-d+2((1+q)+d(1-q))\lambda]\over 4\pi T}+{\cal O}(T^{-2})
 \ee 
 and
\be
v_B(\gamma_{GB})&=&N_\sharp~\sqrt{(1+d)+(1-d)q\over2d}\nn\\
&&- N_\sharp~{k((1+d)+(1-d)q)^{3\over 2}[1-d+2((1+q)+d(1-q))\lambda]\over 32  \pi^2\sqrt{2d}~ T^2}+{\cal O}(T^{-3})\nn\\
 \ee
 Let us make following comments :
 
1. From above results we can see how $\lambda$ and $k$ affect the butterfly velocity.  

2.  At high temperature the butterfly velocities in both sphere and hyperbolic black holes reduce to that in the planar black hole.
 
3. The ratio of the butterfly velocities between that with and without Gauss-Bonnet term is also universal likes as planar case mentioned before.

%%%%%%%%%%
%%%%%%%%%%
\section{Conclusion}

In the first part of this paper we have investigated the butterfly velocity in the anisotropic spacetime. 
We have derived a general formula given in \eqref{mainresult1}.  We used this formula to study  the butterfly velocities in brane systems. 
We have conjectured that the butterfly velocity is invariant under a double-dimensional reduction generally. 
In the second part of the paper we study the butterfly velocity in Einstein-Gauss-Bonnet gravity with arbitrary matter fields. 
The general formula is \eqref{mainresult2}.  
For planar, spherical or hyperbolic black holes the formula reduces to a simple form \eqref{mainresult3}. 
We have considered several examples by using our formula and discussed several interesting properties.
 
 Let us make following final comments:
  
  1.  We have conjectured  that the butterfly velocity is invariant under a double dimensional reduction. The implication of such an invariance is not yet understood and it remains to be clarified.
  
  2. We have found that the ratio of the butterfly velocities between that with and without Gauss-Bonnet term is universal in the sense that it does not depend on the strength of Maxwell field. It would be interesting to understand this result better.

 3. It will be interesting to use \eqref{mainresult2} to investigate systems which are not planar, spherical nor hyperbolic black hole, to see how the Gauss-Bonnet term might affect the butterfly velocity. 

4.  It will be also interesting  find a general butterfly velocity formula in gravity theory with arbitrary combination of quadratic Riemann curvature and extend them to anisotropic space [37].  

Studies along these lines are in progress.

\appendix
\section{Kruskal Coordinate and Useful Relations}
The line element of time and radial parts for a black hole with a horizon at $r=r_H$  is given by
\be
ds^2&=&-a(r)f(r)dt^2+{dr^2\over b(r)f(r)}=-a(r)f(r)\Big[dt^2-dr_*^2\Big]
\ee
where $a(h_h)\ne0, b(h_h)\ne0$ and $f(h_h)=0$. The tortoise coordinate $r_*$ is 
\be
dr_*&=&{dr\over f(r)\sqrt{a(r)b(r)}}
\ee
Near horizon, $f(r)\approx f\rq{}(r_H)(r-r_H)+\cdot\cdot\cdot$, we have a relation
\be
r_*(r)&=&\int_0^r{dr\over f(r)\sqrt{a(r)b(r)}}\approx{1\over  \sqrt{a(r_H)b(r_H)}f\rq{}(r_H)} \ln{r-r_H\over r_H}
\ee
Thus, on the horizon  the tortoise coordinate $r_*(r_H)$ approaches $-\infty$.

The metric can be written in Kruskal coordinate as follows.
\be
ds^2&=& 2A(u v)dudv\\
A(UV)&=&{2a(r)f(r)\over f\rq{}(r_H)^2a(r_H)b(r_H)}e^{-\sqrt{a(r_H)b(r_H)}~f\rq{}(r_H)~r_*}\\
U&=&e^{\sqrt{a(r_H)b(r_H)}{f\rq{}(r_H)\over 2} (-t+r_*)}\\
V&=&e^{\sqrt{a(r_H)b(r_H)}{f\rq{}(r_H)\over 2} (t+r_*)}\\
UV&=& e^{\sqrt{a(r_H)b(r_H)}f'(r_H)~r_*}\\
r_*&=&{1\over \sqrt{a(r_H)b(r_H)}f\rq{}(r_H)}\ln (UV)
\ee
Since  the tortoise coordinate $r_*(r_H)=-\infty$, we see that $U=0$ on the horizon. 

Above definitions and relations imply that
\be
 A(U=0)&=&{2r_H\over f\rq{}(r_H)b(r_H)}
\ee
and 
\be
&&A'(U=0)\equiv {dA(UV)\over d(UV)}\mid_{U=0}={dA(UV)\over d(r_*)}{dr_*\over d(UV)}\mid_{r=r_H}\nn\\
&=&{2~d[a(r)f(r)e^{-\sqrt{a(r_H)b(r_H)}~f\rq{}(r_H)~r_*}]\over f\rq{}(r_H)^2a(r_H)b(r_H)~dr_*}\Big({1\over \sqrt{a(r_H)b(r_H)}f\rq{}(r_H)}~ e^{-\sqrt{a(r_H)b(r_H)}f'(r_H)~r_*}\Big)\mid_{r=r_H}\nn\\
&=&{2r_H^2\over b(r_H)f'(r_H)}~\Big( {a'(r_H) \over a(r_H) }+{f''(r_H)\over f'(r_H)}\Big)
\ee
where  we have used the relation $f(r)\approx f'(r_H)(r-r_H)+{1\over 2}f''(r_H)(r-r_H)^2$. 

Another useful relation  is
\be
h'(U=0)&\equiv&{dh(UV)\over d(UV)}\mid_{U=0}={dh(UV)\over d(r)}{dr\over dr_*} {dr_*\over d(UV)}\mid_{r=r_H}\nn\\
&=&h'(r_H)\Big(\sqrt {a(r)b(r)}~f(r)\Big)\Big({1\over \sqrt{a_hb_h}f'_h} e^{-\sqrt{a_hb_h}f'_h r_*}\Big)\nn\\
&=&r_H~h'(r_H)
\ee

%%%%%%%%%%%%%%%%
%%%%%%%%%%%%%%%%

\section{Derivation of \eqref{old411}}
We first rewrite $(R_{UcUd} R^{cd})^{(1)}$ as
 \be
  (R_{UcUd} R^{cd})^{(1)}&=&  R_{UcUd}^{(0)} R^{(1)cd}+ R_{UcUd}^{(1)} R^{(0)cd}
 \ee
 The first term can be easily calculated using
 \be
 R^{(1)ab}&=&\delta^{aV}\delta^{bV}\Big(-\Big({2A'\over A^3}+{3dh'\over 2A^2h}\Big)\alpha+{1\over Ah}\Delta^{(d)} \alpha\Big)\\
 R_{UVUV}^{(0)}&=&A'
 \ee 
where Laplacian $\Delta^{(d)} \alpha(x)$ is defined in the  d-dimensional metric $ds^2=g_{ij}dx^idx^j$.
We next rewrite the second term as
 \be
R_{UcUd}^{(1)} R^{(0)cd}=R_{UVUV}^{(1)} R^{(0)VV}+R_{UiUj}^{(1)} R^{(0)ij}=R_{UiUj}^{(1)} R^{(0)ij}
 \ee
Note that $R_{UVUV}^{(1)}=0$. 
To calculate the remaining term, $R_{UiUj}^{(1)}$, we write it as
 \be
 R_{UiUj}^{(1)}=(g_{Ua}R^{a}_{~iUj})^{(1)}=g_{UU}^{(1)}(R^{U}_{~iUj})^{(0)}
 +g^{(0)}_{UV}(R^{V}_{~iUj})^{(1)}
 \ee
and 
 \be
g_{UU}^{(1)} (R^{(0)U}_{~iUj})(R^{(0)ij})={h'\over h}\Big[R^{(0)}+{2A'\over A^2}+{dh'\over Ah}\Big]\alpha(x)\delta(U)
\ee
The final step  is to analyze $(R^{V}_{~iUj})^{(1)}$. We find 
\be
(R^{V}_{~iUj})^{(1)}&=&\nabla_U\delta \Gamma^V_{ij}-\nabla_j\delta \Gamma^V_{iU}\nn\\
&=&{1\over 2}\nabla_U\Big[g^{Va}(\nabla_i\delta g_{aj}+\nabla_j\delta g_{ia}-\nabla_a\delta g_{ij})\Big]-{1\over 2}\nabla_j\Big[g^{Va}(\nabla_i\delta g_{aU}+\nabla_U\delta g_{ia}-\nabla_a\delta g_{iU})\Big]\nn\\
&=&-{1\over 2}\nabla_j\nabla_i\Big[g^{Va}\delta g_{aU}\Big]
\ee
Note that, in the second line, the first three terms are zero, while  last two terms are non-zero but they are canceled to each other and therefore only the forth term left.

Using $\delta g_{ab}=-2A(U,V)\alpha(x)\delta(U)\delta_{aU}\delta_{bU}$  we obtain a simple relation
\be
\nabla_j\nabla_i(g^{Va}\delta g_{aU})&=&-2R^{(0)ij}\nabla^{(d)}_i\nabla^{(d)}_j\alpha(x)\delta(U)+\Big[R^{(0)}+{2A'\over A^2}+{dh'\over Ah}\Big]{h'\over Ah}\alpha(x)\delta(U)\nn\\
\ee
where the covariant derivative $\nabla^{(d)}_i$ is defined in a d-dimensional metric, $ds^2=g_{ij}dx^idx^j$.

Collecting the above results we find the relation of $(R_{UcUd} R^{cd})^{(1)}$ given in \eqref{old411}.

%%%%%%%%%%%%%%%%%%%%%%
%%%%%%%%%%%%%%%%%%%%%%
%%%%%%%%%%%%%%%%%%%%%%
\addcontentsline{toc}{section}{References}
\section*{References}
\begin{enumerate}
\item  A. Almheiri, D. Marolf, J. Polchinski, D. Stanford, and J. Sully, ``An Apologia for Firewalls,'' JHEP 09 (2013) 018, [arXiv:1304.6483 [hep-th]].
\item A. Kitaev, ``Hidden correlations in the Hawking radiation and thermal noise,'' (2014), talk given at the Fundamental Physics Prize Symposium, Nov. 10, 2014.
\item D. A. Roberts and D. Stanford, ``Two-dimensional conformal field theory and the butterfly effect,'' Phys. Rev. Lett. 115 (2015) 131603, [arXiv:1412.5123 [hep-th]].
\item J. Maldacena, S. H. Shenker and D. Stanford, ``A bound on chaos,'' [arXiv:1503.01409
[hep-th]].
\item S. H. Shenker and D. Stanford, ``Black holes and the butterfly effect,'' JHEP 1403
(2014) 067 [arXiv:1306.0622 [hep-th]].
\item S. H. Shenker and D. Stanford, ``Multiple Shocks,''JHEP 1412 (2014) 046
[arXiv:1312.3296 [hep-th]].
\item S. Leichenauer, ``Disrupting Entanglement of Black Holes,'' Phys. Rev. D90 (2014) 046009, [[arXiv: 1405.7365 [hep-th]].
\item D. A. Roberts, D. Stanford and L. Susskind, ``Localized shocks,'' JHEP 1503 (2015)
051 [arXiv:1409.8180 [hep-th]].
\item  S. H. Shenker and D. Stanford, ``Stringy effects in scrambling,'' JHEP 1505 (2015) 132
[arXiv:1412.6087 [hep-th]].
\item  D. A. Roberts and B. Swingle, ``Lieb-Robinson and the butterfly effect,'' PRL 117 (2016) 091602 [arXiv:1603.09298 [hep-th]].
\item M. Blake, ``Universal charge diffusion and butterfly effect,'' Phys. Rev. Letter, 117 (2016) 091601,  [arXiv: 1603.08510 [hep-th]].
\item A. P. Reynolds and S. F. Ross, ``Butterflies with rotation and charge,''  [arXiv:1604.04099 [hep-th]].
\item N. Sircar, J. Sonnenschein and W. Tangarife, ``Extending the scope of holographic
mutual information and chaotic behavior,'' JHEP 1605 (2016) 091 [arXiv:1602.07307 [hep-th]]. 
\item W. H. Huang and Y. H. Du, ``Butterfly effect and Holographic Mutual Information
under External Field and Spatial Noncommutativity," JHEP 1702, 032 (2017) [arXiv:1609.08841 [hep-th]].
\item  X. H. Feng, H. Lu, ``Butterfly Velocity Bound and Reverse Isoperimetric Inequality,'' Phys. Rev. D 95, 066001 (2017) [arXiv:1701.05204 [hep-th]]. 
\item  R. G. Cai, X. X. Zeng and H. Q. Zhang, ``Influence of inhomogeneities on holographic mutual information and butterfly effect,'' JHEP 1707 (2017) 082 [arXiv:1704.03989 [hep-th]]. 
\item  V. Jahnke,  ``Delocalizing Entanglement of Anisotropic Black Branes,'' [arXiv:1708.07243 [hep-th]]. 
\item  M. Alishahiha, A. Davody, A. Naseh, S. F. Taghavi,  ``On Butterfly effect in Higher Derivative Gravities,'' JHEP11(2016)032 [arXiv:1610.02890 [hep-th]]. 
\item E. Caceres, M. Sanchez, J. Virrueta,  ``Holographic Entanglement Entropy in Time Dependent Gauss-Bonnet Gravity,''  [arXiv:1512.05666 [hep-th]].
\item M. M. Qaemmaqami,  ``On the Butterfly Effect in 3D gravity,'' [arXiv:1708.07198 [hep-th]]. 
\item Y. Z.  Li, H.  S. Liu, H. Lu, ``Quasi-Topological Ricci Polynomial Gravities,'' [arXiv:1707.00509 [hep-th]]. 
\item  Y. Ling, P. Liu, J. P. Wu,  ``Holographic Butterfly Effect at Quantum Critical Points,'' [arXiv:1610.02669 [hep-th]]. 
\item  Y. Ling, P. Liu, J. P. Wu,  ``Note on the butterfly effect in holographic superconductor models,'' Phys. Lett. B 768 (2017) 288 [arXiv:1610.07146 [hep-th]].
\item  D. Giataganas, U. Gürsoy, J. F. Pedraza,  ``Strongly-coupled anisotropic gauge theories and holography,'' [arXiv:1708.05691 [hep-th]]. 
\item  M. Blake, R. A. Davison, S. Sachdev,  ``Thermal diffusivity and chaos in metals without quasiparticles,'' [arXiv:1705.07896[hep-th]].
\item  D. Ahn, Y. Ahn, H. S. Jeong, K. Y. Kim, W. J. Li and C. Niu, ``Thermal diffusivity and butterfly velocity in anisotropic Q-Lattice models,'' [arXiv:1708.08822[hep-th]].
\item  K. Sfetsos,  ``On gravitational shock waves in curved space-times," Nucl. Phys. B
436, 721 (1995) [hep-th/9408169].
\item  S. Nojiri and S. D. Odintsov, ``Brane-world cosmology in higher derivative gravity or
warped compactification in the next-to-leading order of AdS/CFT correspondence,''
JHEP 07, 049 (2000) [arXiv:hep-th/0006232].
\item T. Dray and G. t'Hooft, ``Gravitational Shock Wave of a Massless Particles", Nucl. Phys. B253 (1985)173.
\item T. Ortin, ``Gravity and String", Cambridge University Press. 2004.
\item D. J. Smith,  ``Intersecting brane solutions in string and M-theory,'' [hep-th/0210157]. 
\item D. G. Boulware and S. Deser,  ``String Generated Gravity Models," Phys. Rev. Lett. 55, 2656 (1985).
\item R. G. Cai,  ``Gauss-Bonnet black holes in AdS spaces," Phys. Rev. D 65, 084014 (2002) [hep-th/0109133].
\item  M. Cvetic, S. Nojiri and S.D. Odintsov,  ``Black Hole Thermodynamics and Negative Entropy in deSitter and Anti-deSitter Einstein-Gauss-Bonnet gravity," Nucl. Phys. B628 (2002) 295,[arXiv : hep-th/0112045].
\item M. B. Gaete and M. Hassaine,  ``Topological black holes for Einstein-Gauss-Bonnet gravity with a nonminimal scalar field," Phys. Rev. D 88 (2013) 104011[arXiv : hep-th/1308.3076].
\item F. Correa and M. Hassaine,  ``Thermodynamics of Lovelock black holes with a nonminimal scalar field,'' JHEP02(2014)014 [arXiv:  hep-th/1312.4516]. 
\item W. H. Huang,  to appear.
\end{enumerate}
\end{document}